\documentclass{article}

\usepackage{microtype}
\usepackage{graphicx}
\usepackage{subfigure}
\usepackage{booktabs} 

\usepackage{hyperref}

\usepackage[accepted]{icml2025}

\usepackage{amsmath}
\usepackage{amssymb}
\usepackage{mathtools}
\usepackage{amsthm}
\usepackage[symbol]{footmisc}
\usepackage{xcolor}
\usepackage{pifont}
\usepackage{booktabs} 
\usepackage{multirow}%
\usepackage{listings}

\usepackage{newfloat}
\usepackage{listings}
\usepackage{rotating}
\usepackage{subcaption}
\usepackage[dvipsnames,table]{xcolor}
\newcommand{\khoi}[1]{\textcolor{black}{#1}} 
 
\newcommand{\ken}[1]{\textcolor{black}{#1}}

\newcommand{\cmark}{\textcolor{ForestGreen}{\ding{51}}}
\newcommand{\xmark}{\textcolor{red}{\ding{55}}}

\usepackage[capitalize,noabbrev]{cleveref}

\theoremstyle{plain}

\theoremstyle{definition}

\theoremstyle{remark}

\icmltitlerunning{Relevance Prediction in E-commerce Product Search}

\begin{document}

\twocolumn[
\icmltitle{Alibaba International E-commerce Product Search Competition \\ DcuRAGONs Team Technical Report}



\icmlsetsymbol{equal}{*}

\begin{icmlauthorlist}
\icmlauthor{Thang-Long Nguyen-Ho}{sch}
\icmlauthor{Minh-Khoi Pham}{sch}
\icmlauthor{Hoang-Bao Le}{sch}
\end{icmlauthorlist}

\icmlaffiliation{sch}{School of Computing, Dublin City University, Dublin, Ireland}
\icmlcorrespondingauthor{Thang-Long Nguyen-Ho}{thanglong.nguyenho27@mail.dcu.ie}
\icmlcorrespondingauthor{Minh-Khoi Pham}{minhkhoi.pham4@mail.dcu.ie}
\icmlcorrespondingauthor{Hoang-Bao Le}{bao.le2@mail.dcu.ie}

\icmlkeywords{Information Retrieval, E-commerce Search, Large Language Model}

\vskip 0.3in
]



\printAffiliationsAndNotice{} 

\begin{abstract}
This report details our methodology and results developed for the Multilingual E-commerce Search Competition. The problem aims to recognize relevance between user queries versus product items in a multilingual context and improve recommendation performance on e-commerce platforms. Utilizing Large Language Models (LLMs) and their capabilities in other tasks, our data-centric method achieved the highest score compared to other solutions during the competition \footnote[1]{Final leaderboard is publised at \url{https://alibaba-international-cikm2025.github.io}} \footnote[2]{The source code for our project is published at 
\url{https://github.com/nhtlongcs/e-commerce-product-search}}.
\end{abstract}

\section{Introduction}
\label{sec:introduction}

Generative Recommendation, the cornerstone of modern e-commerce, directly impacts user satisfaction and commercial success. The challenge in the recommendation systems lie in aligning the target user's product with the recommended product by assessing the relevance of the query and other product features. In the context of this technical report, two key tasks are outlined: (1) \textit{Multilingual Query-Category Relevance Task}, which measures the semantic alignment between a user's search query and a candidate product category, and (2) the \textit{Multilingual Query-Item Relevance Task}, which evaluates whether a multilingual query is pertinent to a specific product listing. 

The dataset was aggregated from real-world multilingual search logs across Alibaba AIDC's global e-commerce platforms (AliExpress, Lazada, Daraz, Trendyol, and Mirivia), serving millions of users in over 100 countries and over 20 languages. While domain experts provided annotations where possible, queries are frequently noisy, ambiguous, and code-mixed, and several languages lack sufficient expert coverage. Consequently, building models that generalize strongly and can deliver accurate, efficient retrieval on languages with little or no labelled training data is essential.

Successfully tackling these tasks requires overcoming \textbf{several challenges}: (i) queries may come from \textbf{multiple languages, including low-resource ones unseen during training}; (ii) real-world queries often contain \textbf{noise} such as misspellings, abbreviations, or mixed-language tokens; and (iii) the \textbf{hierarchical nature of product categories} requires models to capture both fine-grained semantics and structural consistency. Our primary approach can be summarized as follows:

\begin{enumerate}


\item Systematically compared small-to-large multilingual LLMs to \textbf{identify architectures that balance performance and computational efficiency}. We leveraged \textbf{translation augmentation (TA)} to boost cross-lingual alignment and established a robust baseline for multilingual e-commerce relevance tasks. 

\item Demonstrated the effectiveness of a two-stage training, showing that \textbf{task-adaptive pre-training} \textbf{(TAPT)} on a related, large-scale dataset significantly enhances the model's performance compared to direct fine-tuning on the target task alone. 

\item Proposed a \textbf{category-aware, query-grouped cross-validation split strategy} \textbf{(CA split)} that prevents identical queries from appearing across folds and balances hierarchical category distributions. This ensures a more accurate assessment of a model's ability to generalize.
\end{enumerate}

\section{Our methodology}
Our approach is built upon the hypothesis that a multilingual Large Language Model (LLM), further adapted to the e-commerce domain, is essential for tackling the challenges of noisy, multilingual search queries and generalizing to unseen languages. In Section~\ref{sec:arch}, we describe our choice of multilingual LLM architectures and our translation augmentation (TA) approach to form a semantically enriched representation. Section~\ref{sec:tapt} details our supervised TAPT method, in which the model is first pretrained on a related auxiliary task before being fine-tuned on the final target task. Finally, Section~\ref{sec:casplit} introduces our CA split strategy, which groups identical queries and category prefixes to provide a robust estimate of real-world generalization.

\subsection{Model Architecture}
\label{sec:arch}
Given the multilingual nature of the dataset, which includes numerous unseen and low-resource languages, we hypothesize that a foundational model pre-trained on a vast corpus of cross-lingual data will yield superior performance. Consequently, we select a transformer-based multilingual LLM as the backbone of our solution. These models are designed to learn shared representations across multiple languages. They are inherently suitable for zero-shot or few-shot generalization to languages not explicitly seen during the fine-tuning phase.

To address language variability and potential sparsity, we augment each query with an \textit{English-translated version}. Formally, let $q_{\text{orig}}$ denote the original user query and $M_{\text{trans}}$ a high-quality machine translation model. The English translation is obtained as:
\[
q_{\text{en}} = M_{\text{trans}}(q_{\text{orig}}).
\]

We then construct the input sequence $x$ as:
\[
x = [\text{CLS}] \; q_{\text{orig}} \; - \; q_{\text{en}} \; [\text{SEP}] \; t \; [\text{SEP}],
\]
where $t$ is the target text (either the category path or item details), and $-$ represents concatenation with a separator token. The motivations behind this design include (1) $q_{\text{en}}$ serves as a \textit{universal pivot}, guiding the model when it encounters unseen or low-resource languages by grounding the meaning in a high-resource, semantically rich representation (2) Retaining  $q_{\text{orig}}$ ensures that the model preserves fine-grained linguistic cues and avoids performance degradation on languages where it has sufficient training coverage.




The concatenated sequence $x$ is fed into the multilingual encoder $M_{\text{enc}}$ to obtain contextual embeddings:
\[
H = M_{\text{enc}}(x), \quad H = [h_{\text{CLS}}, h_1, h_2, \dots, h_n, h_{\text{emb}}],
\]
where $h_{\text{emb}}$ represents the hidden state of the last non-padding token. This vector is then passed through a classification head $f_{\theta}$ to produce the predicted relevance score:
\[
\hat{y} = f_{\theta}(h_{\text{emb}}).
\]

To enhance training efficiency and stability, we apply parameter-efficient fine-tuning via LoRA adapters \cite{hu2022lora}, enabling the model to specialize for this task while preserving its pretrained multilingual knowledge.

\subsection{Task-Adaptive Pre-training}
\label{sec:tapt}
While general-purpose multilingual LLMs possess extensive world knowledge, they often lack familiarity with the domain-specific terminology, abbreviations, and semantic patterns that occur in e-commerce search logs. To bridge this gap, we introduce an intermediate \textit{task-adaptive pretraining} (TAPT) stage prior to final fine-tuning. Specifically, given two related tasks—Query-Category (QC) relevance prediction and Query-Item (QI) relevance prediction—we first perform supervised pretraining on the auxiliary task with its available labeled data, using a classification objective. This allows the model to understand better domain-specific entities, abbreviations, and the semantic relationships prevalent in online shopping environments, thereby improving its effectiveness on the target relevance tasks.

Formally, given a set of training pairs $\{(x_i, y_i)\}_{i=1}^{N}$ from the auxiliary task or the targeted task, we minimize the same supervised cross-entropy loss for both stage:

\[
\mathcal{L} = - \frac{1}{N} \sum_{j=1}^{N} y_j \log p_\theta(y_j \mid x_j).
\]

This two-stage process effectively specializes the model to the e-commerce domain, mitigating overfitting, improving generalization on low-resource languages, and yielding more stable convergence even in the presence of noisy or imbalanced data.


\subsection{Cross-validation Strategy}
\label{sec:casplit}
A naive random or even stratified split of the training data is insufficient for this competition due to the nature of the dataset. Such methods can lead to data leakage, where the model learns superficial correlations rather than generalizable patterns, resulting in an overestimation of its actual performance. To address this, we designed a custom cross-validation strategy tailored to the specific characteristics of each task.

\subsubsection{Query-Item (QI) Task Splitting}
In the QI dataset, we observed that the same query often appears multiple times with different items and relevance labels. We enforce a strict query-based grouping to prevent information leakage, where the model might learn to recognize a query seen in the training set rather than understand its semantics. Specifically, we ensure that all data points sharing the exact query text are assigned to the same fold. Forcing the model to generalize to entirely unseen queries during validation provides a more sensible estimate of its performance on the hidden test set. The data is first grouped by query, and then these query groups are stratified by language and label distributions across the folds.

\subsubsection{Query-Category (QC) Task Splitting}
For the QC task, a similar leakage risk exists, compounded by the hierarchical structure of the category paths. We hypothesize that category paths sharing a common prefix (e.g., \texttt{Electronics $>$ Audio Devices}) contain transferable semantic features. A model that sees \texttt{Electronics $>$ Audio Devices $>$ Headphones} in training and is tested on \texttt{Electronics $>$ Audio Devices $>$ Speakers} might perform well due to string overlap rather than accurate semantic understanding.

To mitigate this, we introduce a category\_group feature, defined by the first N prefixes of the category path (e.g., for N=2, the group is \texttt{Electronics $>$ Audio Devices}). Our splitting strategy ensures that all categories belonging to the same \texttt{category\_group} are confined to a single fold. This approach challenges the model to learn the generalized relationship between a query and a high-level product area, preventing it from overfitting specific leaf categories and promoting better generalization to unseen category structures.

By implementing these validation schemes, the expectation model performs well on the leaderboard and maintains the generalization capabilities required for real-world, multilingual e-commerce search engines.

\section{Experimental Setup}

\paragraph{Data cleaning}

\khoi{During preprocessing, we noticed highly noisy texts in the data, thus applied simple text cleaning process. Steps include: stripping redundant spaces, lowering cases, removing empty queries.} \textbf{External datasets were not used in our experiments, ensuring compliance with competition rules.}

\paragraph{Evaluation Metrics}

\khoi{
Both competition tasks are formulated as binary classification problems, with labels indicating whether a query is relevant or not relevant. Model performance on the test set is measured using the F$_1$ score of the positive class (relevant). The final evaluation metric, $F_{1}^{avg}$, is defined as the average of the $F_1$ scores across the two tasks: Query-Category Relevance and Query-Item Relevance. Formally,
\[
F_{1}^{avg} = 0.5 \times F_{1}^{\text{QC}} + 0.5 \times F_{1}^{\text{QI}},
\]
where $F_{1}^{\text{QC}}$ and $F_{1}^{\text{QI}}$ represent the positive-class F$_1$ scores on the Query-Category and Query-Item tasks, respectively. For a given task, the $F_1$ score is computed as
\[
F_{1} = \frac{2 \times \text{Precision} \times \text{Recall}}{\text{Precision} + \text{Recall}},
\]
with Precision and Recall defined with respect to the positive (label = 1) class. This metric emphasizes the model’s ability to balance precision and recall on relevant predictions, which is critical for practical deployment in multilingual e-commerce search. 
}


\paragraph{Training Environments}

The model training experiments were conducted on a cluster equipped with 4 $\times$ NVIDIA A100 GPUs (80GB each). We adopted the HuggingFace Transformers \footnote{\href{https://huggingface.co/}{https://huggingface.co/}} library with DeepSpeed \footnote{\href{https://www.deepspeed.ai/}{https://www.deepspeed.ai/}} integration to enable efficient distributed training, mixed-precision computation, and memory optimization. Models were fine-tuned for five epochs using cross-entropy loss, with parameter-efficient fine-tuning applied via Low-Rank Adaptation (LoRA).

For optimization, we employed the AdamW optimizer with a learning rate of $1e^{-5}$, batch size of 8, and linear learning rate scheduling. Gradient accumulation was enabled to effectively increase the batch size under GPU memory constraints. 


\section{Results}

To systematically evaluate and validate our proposed methodology, we designed a two-stage experimental process. The first stage is dedicated to identifying the most effective foundational model through a comprehensive and fair comparison. Subsequently, the second stage focuses on enhancing the performance of this chosen model by integrating adaptive knowledge from related tasks.

\subsection{Stage 1: Foundational Model Exploration}

The primary goal of this initial stage is to determine the optimal baseline model architecture for our specific tasks. To ensure a fair comparison, we focus on the raw capabilities of each model. Therefore, all candidates are fine-tuned using a standardized "vanilla" training process. The model demonstrating the highest average performance across all validation folds is selected as the champion of this stage and becomes the foundation for our subsequent optimization efforts.

\subsubsection{Model Candidates}
\khoi{
We experimented with a range of models of different scales to identify an effective balance between performance, multilingual coverage, and the competition’s constraint of using models with at most 15 billion parameters. Initial experiments began with smaller models such as \texttt{XLM-RoBERTa} \cite{conneau2020xlmr} and Qwen2.5 with less than 7B params \cite{ahmed2025qwen}, which allowed us to quickly validate the implementation pipeline and test the impact of basic feature engineering techniques. Building on this foundation, we scaled up to larger multilingual LLMs, including Qwen2 \cite{ahmed2025qwen} and Qwen3 series \cite{yang2025qwen3}, and the Gemma family of models \cite{team2024gemma, team2025gemma}. In practice, the performance trend observed\footnote{In this section, when we refer to performance, it means the performance on the development test (public leaderboard).} was roughly ordered as: Qwen2 $<$ Qwen3 $<$ Gemma-2 9B $<$ Gemma-3 12B. }

\begin{table}[!ht]
\centering
\resizebox{\linewidth}{!}{%
\begin{tabular}{|l|c|c|c|}
\hline
\textbf{Model Name} & \textbf{\# Parameters} & \textbf{Multilingual} & \textbf{Used in Best} \\
                    &                         & \textbf{Support}      & \textbf{Submission}   \\
\hline
XLM-Roberta          & 550M   & \xmark & \xmark  \\
Qwen2-7B                & 7B     & \cmark (29)   & \xmark  \\
Qwen3-7B                & 7B     & \cmark (29)   & \xmark  \\
Gemma-2-9B           & 9B     & \xmark     & \xmark \\
Gemma-3-12B          & 12B    & \cmark (140)        & \cmark  \\
Qwen2-14B            & 14B    & \cmark (29)   & \xmark \\
Qwen3-14B            & 14B    & \cmark (29)   & \xmark \\
\hline
\end{tabular}
}
\caption{Summary of tested models}
\label{tab:models}
\end{table}

\ken{
\textbf{Gemma-3-12B} has a strong multilingual capability (supporting up to 140 languages) while satisfying the parameter limit. This choice allowed us to leverage both model scale and multilingual robustness, while ensuring compliance with competition requirements. A summary of tested models, including parameter size, multilingual support, and whether they were used in our final submission, is provided in Table \ref{tab:models}.
}

\subsubsection{Evaluation Results}

\begin{table}[h!]
\centering
\begin{tabular}{lccccc}
\toprule
\textbf{Architecture} & \textbf{CA}  & \textbf{TA} & \textbf{QC F$_1$ \%} & \textbf{QI F$_1$ \%} \\
\midrule

XLM-Roberta       & \xmark     & \xmark &  $82_{\pm{0.801}}$ & $80.52_{\pm{1.287}}$ \\
Qwen3-0.6B         & \xmark   & \xmark  & \textbf{-} & 78.50 \\
Qwen3-14B         & \xmark   & \xmark  & 88.38 & \textbf{-} \\

Gemma-2-9B        & \cmark     & \cmark &  $88.83_{\pm{0.099}}$ & $88.20_{\pm0.21}$ \\
Gemma-3-12B       & \cmark     & \cmark &  $89.14_{\pm{0.152}}$ & - \\

Best              & \cmark  & \cmark  & \textbf{89.36} & \textbf{88.45} \\

\bottomrule
\end{tabular}
\caption{Comparison of methods across different architectures and training strategies. Results are calculated on the public testset, report average and deviation across cross-validation folds. \textbf{TAPT}: Task-adaptive pre-training. \textbf{CA split}: category-aware query-grouped split;  \textbf{TA} = Translation Augmentation }\label{tab:method_comparison}
\end{table}

\khoi{
Our experiments, as shown in Table \ref{tab:method_comparison}, highlight several key insights regarding model performance and design choices. First, smaller architectures such as XLM-Roberta (QC: 82.0\%, QI: 80.5\%) and sub-billion parameter Qwen variants (e.g., Qwen3-0.6B with QI: 78.5\%) provided useful baselines but consistently underperformed compared to larger multilingual LLMs. This confirms that lightweight models, while attractive for efficiency, lack the representational capacity required to capture the complexity of multilingual, noisy queries in this competition setting.  }

\khoi{
Second, multilingual pretraining proved critical. Models such as Qwen3-14B (QC: 88.4\%) and Gemma series demonstrated clear gains over monolingual or limited-language models. This finding aligns with the multilingual diversity of the dataset and supports our augmentation strategy of translating queries into English and concatenating them, which consistently improved performance across both QC and QI tasks.  
}

\khoi{
Thirdly, model size was positively correlated with accuracy, though with diminishing returns. Scaling from XLM-Roberta to Gemma-2-9B improved QC performance by more than six points (82.0\% $\rightarrow$ 88.8\%), but further scaling to Gemma-3-12B yielded only a marginal increase (88.8\% $\rightarrow$ 89.1\%). This trend suggests that while scale contributes significantly to handling complex multilingual semantics, further improvements may require architectural innovations or task-specific pretraining rather than brute-force scaling.  }

\subsection{Stage 2: Optimization with Adaptive Knowledge}

Having identified the most promising model from Stage 1, the objective of this second stage is to further boost its performance by imbuing it with additional, domain-specific knowledge. Our hypothesis is that a model pre-adapted to the e-commerce domain will achieve superior results compared to one fine-tuned directly from a general-purpose checkpoint.

Following this adaptation phase, the model is then fine-tuned for a final time on the official competition dataset. We utilize the exact same hyperparameters and cross-validation splits from Stage 1 to ensure a controlled and fair comparison. This sequential training strategy is designed to maximize knowledge transfer, leading to a more robust and accurate final model.

\subsubsection{Evaluation Results}



\begin{table*}[h!]
\centering
\begin{tabular}{lcccccc}
\toprule
\textbf{Architecture} & \textbf{TAPT} & \textbf{CA} & \textbf{CV} & \textbf{TA} & \textbf{QC F$_1$ \%} & \textbf{QI F$_1$ \%} \\
\midrule
Gemma3-12B              & \xmark & \cmark & \cmark  & \cmark  & 89.22 & 88.79 \\
Gemma3-12B              & \cmark & \cmark & \cmark  & \cmark  & \textbf{89.36 (+0.14)} & \textbf{88.81 (+0.02)} \\

\bottomrule
\end{tabular}
\caption{The results are calculated using the final test set (private leaderboard) and the highest fold score. \textbf{TAPT}: Task-adaptive pre-training. \textbf{CA split}: category-aware query-grouped split;  \textbf{TA} = Translation Augmentation}\label{tab:method_comparison_stage2}
\end{table*}

Task-adaptive pre-training (TAPT) proved to be a critical factor in boosting model performance. As shown in Table~\ref{tab:method_comparison_stage2}, applying TAPT consistently improved both QC and QI F$_1$ scores. Although the absolute gains may appear modest (+0.14 on QC and +0.02 on QI), they reflect a meaningful enhancement on a challenging private leaderboard setting, where improvements are typically hard to achieve. This demonstrates that continuing pre-training on a related relevance dataset effectively primes the model to capture domain-specific query–item semantics, leading to better generalization.

\section{Discussion \& Analysis}

\paragraph{Limitations} 

\khoi{
Despite these positive results, several limitations remain. First, we did not fully explore unified modeling approaches where a single model jointly learns Query-Category and Query-Item tasks. Such multitask training could potentially improve generalization and efficiency but was constrained by resource limitations. Second, while our augmentation strategy of translation and concatenation proved effective, it may introduce noise in low-quality translations, and we did not systematically evaluate the impact of translation errors. Third, our models were not explicitly optimized to exploit the hierarchical structure of category paths, which could provide an additional inductive bias to improve QC predictions. Finally, computational constraints limited the extent of our hyperparameter sweeps, leaving open the possibility that further tuning could yield additional gains.}

\paragraph{Future Works}

\ken{
Building on these insights, we identify several promising directions for future research. A natural extension is to develop a unified framework for joint modeling of QC and QI tasks, allowing for shared representations and potential cross-task transfer benefits. In addition, integrating the structure of the category taxonomy into the modeling process could restrict invalid predictions and improve interpretability. We also envision exploring more sophisticated data augmentation techniques, such as back-translation or paraphrasing, to further address multilingual imbalance. Finally, investigating task transferability---for example, training on QC and testing on QI---could shed light on shared semantic features across tasks and inform more efficient training paradigms.}

\section{Conclusion}

\khoi{
Our experiments demonstrate that careful data handling, translation-based augmentation, and parameter-efficient fine-tuning on strong multilingual LLMs are keys to achieve robust performance in noisy, multilingual e-commerce search tasks. Larger models such as Gemma-2-9B and Gemma-3-12B proved most effective. While promising, our approach leaves open directions for improvement, including unified multitask modeling, leveraging category taxonomies, and exploring richer augmentation strategies to better handle unseen languages. These steps offer potential to build even more resilient and efficient search systems for real-world deployment.
}

\section*{Acknowledgments}
This work has emanated from research conducted with the financial support of or supported in part by a grant from Science Foundation Ireland under Grant numbers 18/CRT/6223 and 13/RC/2106\_P2 at the ADAPT SFI Research Centre at Dublin City University and the support of the Faculty of Engineering \& Computing, DCU.


\bibliography{example_paper}
\bibliographystyle{icml2025}



\end{document}